\documentclass[12pt]{article}
\usepackage{epsf}
\usepackage{pazh}
\usepackage{flushrt}

\begin{document}

{\tiny To be published in Astronomy Letters, Vol. 27, No. 12, 2001, pp.
790-793\\ Translated from Pis'ma v Astronomicheskij Zhurnal, Vol. 27, No.
12, 2001, pp. 919-922}

\title{Correlation between Break Frequency and Power Density Spectrum Slope
for the X-ray Source Cygnus~X-2: RXTE/PCA Data}{Cyg X-2: Correlation between
Break Frequency and PDS Slope.}

\author{Sergey~Kuznetsov$^{1,2}$}
\affil{\it $^{1}$ Space Research Institute, Profsoyuznaya 84/32, Moscow,
117810 Russia\nl 
$^{2}$ CEA, DSM/DAPNIA/SAp, Centre d'Etudes de Saclay, 91191 Gif-sur-Yvette
Cedex, France}

\begin{center}
{\small Received: June 25, 2001\\}
\end{center}

\section*{}
{\small We present results of RXTE observations of the X-ray source
Cyg~X-2 during 1996--1999. Its power density spectra in the
$0.1$--128-Hz band are fitted by a model that takes into account the
power-law spectral behavior at frequencies below and above the break
frequency, with an introduction of one or more Lorenz lines to
describe the peaks of quasi-periodic oscillations that correspond to
the horizontal branch of the Z~track. The RXTE observations have
revealed a positive correlation between the break frequency and the
indices of the two parts of the spectrum. The spectrum steepens with
increasing break frequency both above and below the break frequency.}
\clearpage

\section{Introduction}
Cygnus~X-2 belongs to low-mass binaries with accreting neutron stars
and is one of the brightest X-ray sources. By its spectral
characteristics, Cyg~X-2 belongs to the class of Z-type sources
(Hasinger and van der Klis 1989), which are characterized by a
Z-shaped track in the color-color diagram (CCD). In this
interpretation, the spectral properties are presented in the ``hard''
and ``soft'' colors, each of which is the harder-to-softer flux ratio
in the corresponding energy band. The Z-shaped track is commonly
divided into three parts called branches: the horizontal (HB, the
upper part of the diagram), normal (NB, the intermediate part), and
flaring (FB, the lower part) branches. The position along the Z~track
is generally believed to be associated with increase in mass
accreation rate of from HB to FB. Six sources are currently known to
exhibit Z~tracks in the color-color diagram: Scorpius~X-1, Cygnus~X-2,
GX~17+2, GX~5-1, GX~340+0 and GX~349+2.


The power density spectra (Fourier transforms of the flux) of Z-type sources
exhibit low-frequency (5--100~Hz) quasi-periodic oscillations (QPOs) of the
X-ray flux. The names of the QPOs correspond to the branch with which their
origin is identified: horizontal- (HBO), normal- (NBO), and flaring-branch
(FBO) oscillations. HBOs (15--100~Hz) can also be detected in the NB
spectral state. However, as one recedes from HB, the significance of the QPO
peaks decreases, and they become undetectable. When moving along the Z~track
in its NB--FB segment, a QPO peak in the range 5--20~HZ (NBO/FBO) emerges in
the power density spectra.

All three types of QPOs characteristic of the low-frequency part ($<100$~Hz)
of the power density spectrum have been detected for the source Cyg~X-2. The
NBO and FBO frequencies are very close to the break frequency, which
introduces a large uncertainty in its determination.  For this reason, we
excluded from our analysis those observations in which NBO/FBOs were
detected.

\section{Data and observations}

For our time analysis, we used the archival data of the PCA (Proportional
Counter Array) instrument (Jahoda et al. 1996) onboard the RXTE
observatory (Bradt et al. 1993) retrieved from the Goddard Space
Flight Center Electronic Archive.

The X-ray source Cyg~X-2 was observed by the RXTE observatory during nine
series of pointing observations (10063, 10065, 10066, 10067, 20053, 20057,
30046, 30418, 40017): in March, August, October 1996, June, July,
September 1997, July 1998, and in separate sessions from July until October
1998 and from January until August 1999. The observations of Cyg~X-2 over
this period correspond to three different observational epochs of RXTE/PCA
(2, 3, and 4 in the adopted classification), for which the boundaries of the
PCA energy channels were changed.

To construct the power density spectra, we used observations with a
resolution of $\sim$122 $\mu$s ($2^{-13}$ s) from the 14th to 249th
PCA energy channels.  This range corresponds to the flux of detectable
photons up to $\sim$60~keV, whose lower limit begins from
$\sim$4.3~keV, $\sim$5.0--5.3~keV, and $\sim$5.8~keV for epochs 2, 3,
and 4, respectively.  In this energy band, the detection of QPOs
corresponding to the horizontal branch of the Z~track is most
significant. The power density spectra were obtained by the standard
method of Fast Fourier Transform (van der Klis 1989).

We combined the observational data that were not represented by a single
format for all channels from 14th to 249th. Of all the observations, we used
only those during which the angle between the source direction and the
Earth's horizon was more than 10$^{\circ}$ and the PCA axis was offset from
the target by no more than 0.02$^{\circ}$.  Among the observations of
Cyg~X-2, all five proportional counters were not always be switched on to
record events. If the operating condition of one of the counters changed
during a continuous observation (whose duration did not exceed the duration
of one orbit and was, on the average, $3-3.5\times10^3$~s), then the time
interval during which the total count rate changed abruptly was excluded
from the analysis. Because of this filtering, the total usable observational
time for Cyg~X-2 was more than $4\times10^5$~s.

To analyze the low-frequency ($<100$~Hz) variability of Cyg~X-2, we
constructed power density spectra in the range $0.03125-128$~Hz. No
corrections were made for background radiation and for dead time
(attributable to the instrumental delay in recording events).

\section{Results}

Fitting the power density spectra by a constant and by a power-law at
frequencies below and above the break frequency did not yield acceptable
results (according to the $\chi^{2}$ test).  The main reason was the absence
of a sharp break and the uncertainty in the measurement of its position in
the power density spectrum. The model in which at frequencies much higher
($\nu/\nu_{\mathrm{break}} \gg 1$) and much lower ($\nu/\nu_{\mathrm{break}}
\ll 1$) than the break, each part of the spectrum could be fitted by its own
power-law and in which the transition between them was not jumplike proved
to be more suitable:
\begin{eqnarray}
P =
C\frac{\nu^{-\alpha}}{1+\left(\frac{\nu}{\nu_{break}}\right)^{\beta}}.
\end{eqnarray}

Thus, $P \propto \nu^{-\alpha}$ at $\nu/\nu_{\mathrm{break}} \ll 1$ and $P
\propto \nu^{-\alpha-\beta}$ at $\nu/\nu_{\mathrm{break}} \gg 1$.

The power density spectra were fitted in the $0.1-128$-Hz band by this
model with the additional introduction of one or two Lorenz lines to
allow for the peaks of QPOs and their harmonics. To take into account
the PCA dead-time effect, which causes the total level to be shifted
to the negative region (because of this effect, the Poissonian noise
level subtracted from all spectra differs from $2.0$ in Leahy
normalization units; see van der Klis (1995) for more details), we
added a constant to the general model.


Figure~1 shows typical power density spectra of Cyg~X-2 for various measured
break frequencies $\nu_{\mathrm{break}}$. The upper spectrum was constructed
from the observations on August 31, 1996 (7:04--8:00 UTC) and has the
following best-fit parameters: $\nu_{\mathrm{break}}=3.1\pm0.3$,
$\alpha=-0.13\pm0.04$, $\beta=1.45\pm0.03$,
$\nu_{\mathrm{Lorenz}}=20.17\pm0.05$, $\nu_{\mathrm{2Lorenz}}=39.0\pm0.4$,
$\chi^{2}=236$ (217 degrees of freedom). The lower power density spectrum
was obtained on March 24, 1996 (2:27--3:19 UTC), which was scaled by a
factor of $0.01$ and has the following parameters:
$\nu_{\mathrm{break}}=12\pm1$, $\alpha=0.30\pm0.03$, $\beta=2.0\pm0.2$,
$\nu_{\mathrm{Lorenz}}=45.0\pm0.7$, $\chi^{2}=242$ (220 degrees of freedom).
In Fig.~1, we clearly see a difference between the two spectra. The best
fits to each of the spectra (solid curves) are shown on the corresponding
scale.

For all the selected data, we obtained satisfactory best-fit
parameters. The break frequency turned out to positively correlate
with the indices for the low-frequency and high-frequency parts of the
power density spectrum ($0.1-128$~Hz).  In Fig.~2, the indices
$\alpha$ and $\beta$ of model~(1) are plotted against break frequency
$\nu_{\mathrm{break}}$, although in reality, the power-law spectral
slope for $\nu/\nu_{\mathrm{break}} \gg 1$ tends to $-\alpha-\beta$,
and the correlation is preserved. The open circles in Fig.~2 indicate
the data whose power density spectra exhibit two QPO peaks. The ratio
of the peak frequencies is close to 2. As the break frequency
increased, the significance of the HBO peaks reduced. The filled
circles indicate the data in which only the main QPO peak was detected
and a second harmonic (probably) of the main peak was either
undetectable or its significance was at a confidence level lower than
$3\sigma$.

The data in Fig.~2 were fitted by straight lines. For each of the indices,
we derived the following parameters: $\alpha\approx
-0.18+0.04\nu_{\mathrm{break}}; \beta\approx 1.23+0.06\nu_{\mathrm{break}}$.

\section{Discussion}

For the Z-type sources (to which Cyg~X-2 belongs), the typical
power-law index for the part of the spectrum above the break frequency
lies within the range $\sim 1.5-2.0$ (van der Klis 1995). In papers on
a time analysis of the low-frequency part of the power density
spectrum \mbox{($< 100$~Hz)} for Cyg~X-2 (Kuulkers 1999), the
variability of the source below the break frequency is assumed to be
constant and is fitted by a constant. We see from Fig.~2 that the
power-law slopes are equal to their assumed values for the spectra
with break frequencies below $\sim$ 10~Hz. This range of break
frequencies roughly corresponds to the position of the source on the
horizontal branch of the Z~track. For the other spectral states (NB
and FB), the break frequency is difficult to determine because of the
emergence of NBOs/FBOs at close frequencies or because of the absence
of a visible break in the power density spectrum (power-law shape of PDS
in the range $0.1-100$~Hz).

Here, we analyzed all the available RXTE observations of Cyg~X-2. The
correlation between the power-law indices below and above the break
frequency has been found for the first time. It could be anticipated
that a similar correlation is a common property of the Z-type sources.

\section*{Acknowledgements}

This study was supported by a French ``GDR PCHE'' grant, which made visit of
author to Service d'Astrophysique possible. The work was supported in part
by the Russian Foundation for Basic Research (project No.~00-15-96649). This
research has made use of data obtained through the High Energy Astrophysics
Science Archive Research Center (HEASARC), provided by the NASA/Goddard
Space Flight Center. Author is grateful to L.~Titarchuk, B.~Stone and
Ph.~Laurent and also would like to thank N.~White, G.~Swank, F.~Newman, and
J.~Repaci for the opportunity to work with the RXTE archival data on compact
media. S.Kuznetsov is grateful to M.~Revnivtsev for providing valuable
remarks.

\clearpage
\section*{References}

{Bradt H., Rotschild R., Swank J., Astron. Astrophys., Suppl. Ser. 97,
335 (1993).} 

{Hasinger G., van der Klis M., Astron. Astrophys. 225, 79 (1989).}

{Jahoda K., Swank J., Giles A., Stark M., Strohmayer T., Zhankg W.,
Morgan E., Proc. SPIE 2808, 59 (1996).}

{Kuulkers E., Wijnands R., van der Klis M., Mon. Not. R. Astron. Soc. 308,
485 (1999)}

{van der Klis M., in {\it Timing Neutron Stars}, Ed. by \"Ogelman H., van
der Heuvel E.P.J. (Kluwer, Dordrecht, 1989) NATO ASI Ser., Vol.~360, p.~27.}

{van der Klis M., in {\it Lives of the the Neutron Stars}, Ed. by 
Alpar M., Kizilo\v{g}lu \"U., Van Paradijs J. (Kluwer, Dordrecht, 1995),
NATO ASI Ser., Vol.~450, p.~301.}

\newpage


\begin{figure}
\figurenum{1}
\plotone{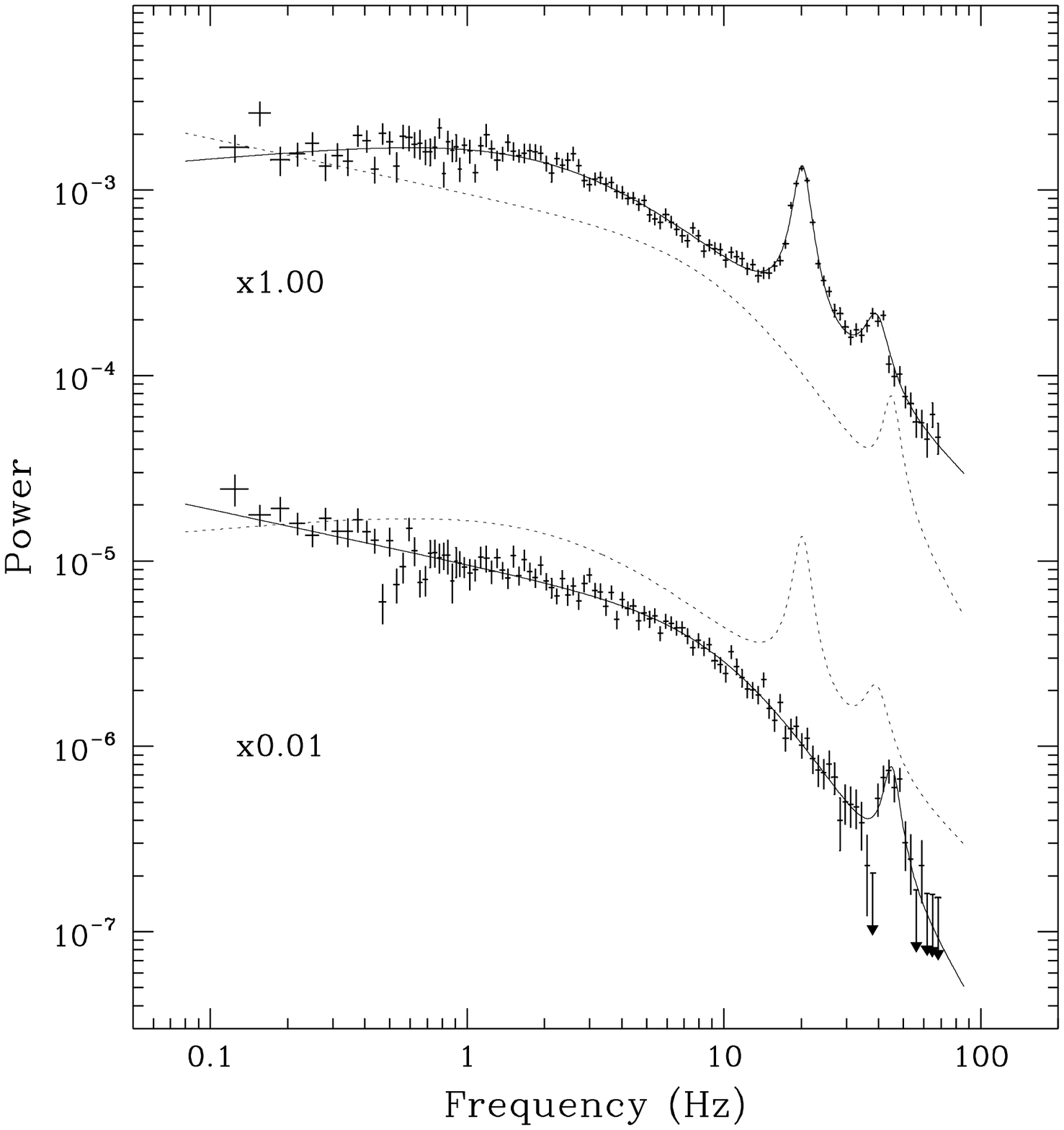}
\epsscale{1.0}
\caption{Two power density spectra for Cyg~X-2 in the 5--60-kev energy band.
The power of the lower spectrum was reduced by a factor of 100. The spectra
are fitted by the break model and by Lorenz lines (solid curves); the dotted
lines represent scalable fits to each of the spectra. The peaks in the upper
and lower spectra correspond to the first and second harmonics of HBOs and
only to its first harmonic, respectively.}
\end{figure}

\begin{figure}
\figurenum{2}
\plotone{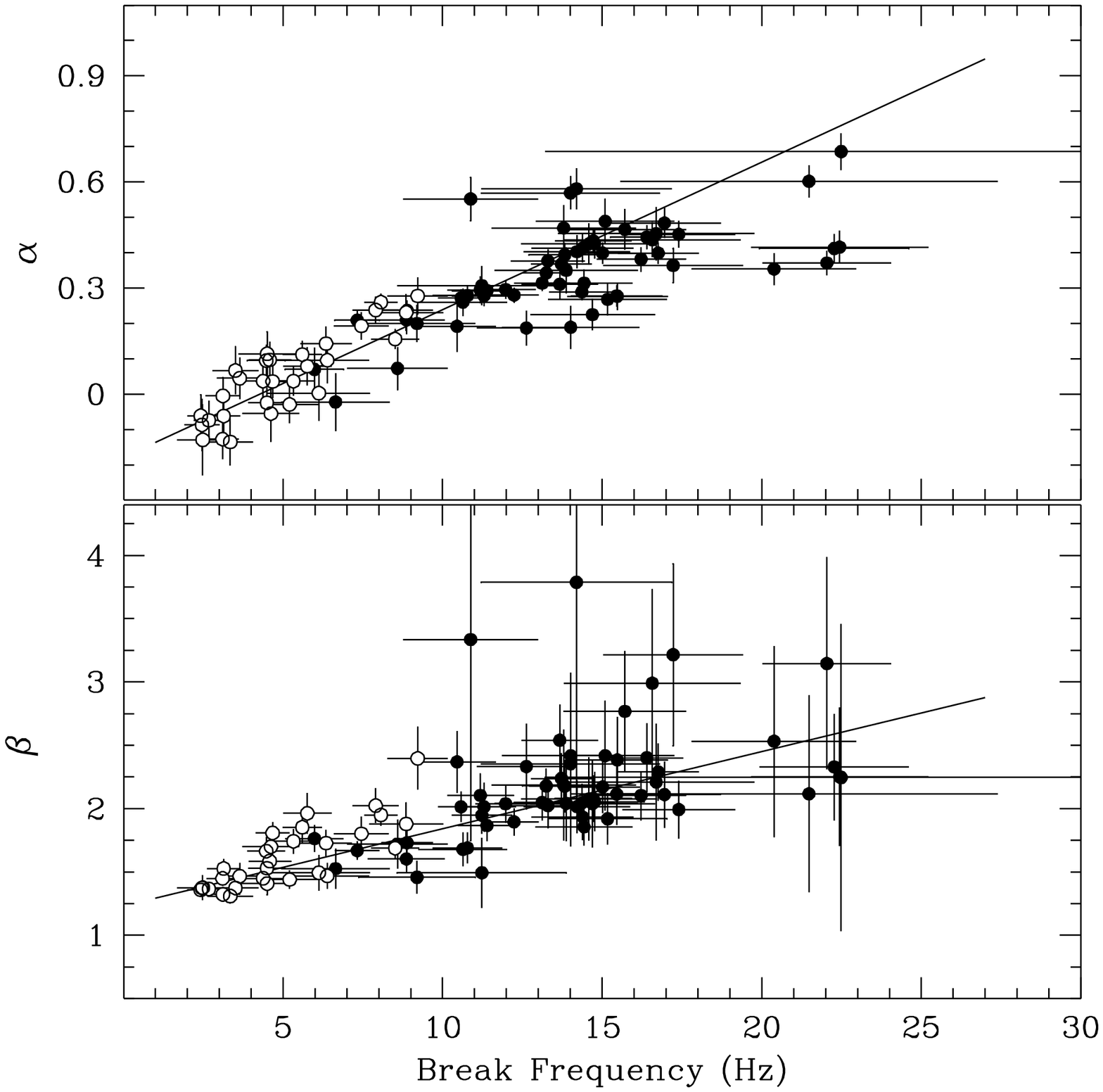}
\epsscale{1.0}
\caption{Model indices $\alpha$ and $\beta$ versus break frequency. In the
upper panel, the index corresponds to the fitting range at frequencies below
the break; the lower panel shows an additional index that is introduced to
fit a steepening of the spectrum above the break frequency. The best-fit
parameters for those power density spectra that, apart from the fundamental
HBO harmonic, contain the second harmonic are indicated by open circles. The
solid line represents a straight-line fit to the data. For the data shown in
the upper and lower panels, the straight line was drawn by taking into
account errors in the break frequency and in the index $\beta$,
respectively.}

\end{figure}

\end{document}